\documentclass[a4paper, preprint]{article}
\usepackage{graphicx}
\usepackage{setspace}
\begin{document}
\title{Isoelectronic series of oxygen deficient centers in silica: experimental estimation of homogeneous and inhomogeneous spectral widths}
\author{Michele D'Amico$^{1,2,*}$, Fabrizio Messina$^{1}$,\\ Marco Cannas$^{1}$, Maurizio Leone$^{1,2}$, and Roberto Boscaino$^{1}$}
\maketitle

$^1$Dipartimento di Scienze Fisiche ed Astronomiche,
Universit$\grave{a}$ di Palermo, Via Archirafi 36, I-90123 Palermo, Italy\\

$^2$Istituto di Biofisica, U.O. di Palermo, Consiglio Nazionale delle Ricerche, Palermo, Italy\\

$^*$ Correspondent author: damico@fisica.unipa.it
\paragraph{Abstract}
We report nanosecond time-resolved photoluminescence measurements on
the isoelectronic series of oxygen deficient centers in amorphous
silica, Si-ODC(II), Ge-ODC(II) and Sn-ODC(II) which are responsible
of fluorescence activities at $\sim$4 eV under excitation at $\sim$5
eV. The dependence of the first moment of their emission band on
time, and that of the radiative decay lifetime on emission energy
are analyzed within a theoretical model able to describe the effects
introduced by disorder on the optical properties of the defects. We
obtain separate estimates of the homogeneous and inhomogeneous
contributions to the measured emission linewidth and we derive
homogeneous spectroscopic features of the investigated point defects
(Huangh-Rhys factor, homogeneous width, oscillator strength,
vibrational frequency). The results point to a picture in which an
oxygen deficient center localized on a heavier atom features a
higher degree of inhomogeneity due to stronger local distortion of
the surrounding matrix. For Si, Ge, Sn related defects the parameter
$\lambda$, able to quantify inhomogeneity, results to be 65, 78 and
90$\%$, respectively.

\section{Introduction}
Amorphous silicon dioxide (silica) is a wide band-gap insulator
widely used in present optical and electronic technologies. As a
matter of fact, it is the material of choice to fabricate the
insulating films incorporated in common metal-oxide-semiconductor
(MOS) transistors, and optical components to be used in the
ultraviolet (UV) spectral range, such as lenses, fibers, and Bragg
gratings.\cite{Erice, nalwa} Since the presence of point defects in
SiO$_2$ significantly compromises the performance of the material in
applications, defects in SiO$_2$ are an important and widely debated
technological problem in current literature.

On the other side, the physics of color centers embedded in an
amorphous matrix is a fundamental scientific problem which poses
several unanswered questions.\cite{stoneham} In a crystal, members
of an ensemble of identical defects are virtually indistinguishable
from the spectroscopical point of view, since their local
environments are identical as well due to translational invariance
of the solid. Hence, absorption or photoluminescence (PL)
lineshapes, as well as decay lifetimes, are to be considered in this
case as homogeneous properties of the defects. On the other hand,
defects in an amorphous solid are supposedly characterized by
statistical distributions of the spectroscopic properties, since the
disorder of the matrix gives rise to site-to-site differences among
the environments experienced by the single centers. For this reason,
it is generally assumed that the lineshapes of their optical bands
feature an \emph{inhomogeneous broadening}, \cite{Erice, nalwa,
holeburning} so that the overall spectroscopic properties of an
"amorphous defect" are determined by the concurrence of homogeneous
and inhomogeneous effects.

While this general interpretation scheme is widely accepted in
literature, no general approach is currently available to separate
the homogeneous and inhomogeneous contributions to the
experimentally observed linewidth of an absorption or PL band.
Moreover, almost nothing is known on the effect of inhomogeneity on
the decay properties of a defect. The homogeneous absorption
linewidth is due to electron-phonon interaction, and is determined
by basic physical properties of the defect, namely the Huangh-Rhys
factors and the phonon vibrational frequencies.\cite{Erice, nalwa}
In the past, many techniques were used to attempt its estimation in
several heterogenous systems: exciton resonant luminescence,
resonant second harmonic scattering, femtosecond photon echo,
spectral hole burning.\cite{furumiya, kuroda, mittlemann, woggon}
This last technique is particularly interesting also for its
technological implications, namely the promising feature of writing
information bits as homogeneous "holes" in a sufficiently broad
inhomogeneous optical band, thus creating solid state optical
memories with very high bit density. \cite{nalwa, holeburning}

In a recent paper,\cite{damicoPRB} we have proposed a new
experimental approach to this problem, which allows to estimate the
inhomogeneous and homogeneous linewidths of a defect based on
studying the variations of the radiative decay lifetime within an
inhomogeneously broadened emission band. Our aim here is to apply
this technique to a particular kind of point defect in silica, the
so-named Oxygen Deficient Center of the second type, shortly
ODC(II). The ODC(II) exists as an intrinsic defect or in two
extrinsic varieties related to the impurity content. Its mostly
accepted structural model consists in a twofold coordinated atom
(=X$^{\bullet \bullet}$)\cite{skuja1984, nishikawa1992,
skuja1994}, where X can be either a Si (Si-ODC(II)), a Ge
(Ge-ODC(II)) or a Sn (Sn-ODC(II)) atom. However, the structure of
Si-ODC(II) is still debated at the moment \cite{SkujaReview98}, as
it has been proposed an alternative model of the defect as a neutral
oxygen vacancy between two silicon atoms
($\equiv$Si-Si$\equiv$).\cite{jones, imai} Due to their
isoelectronic outer valence shell (ns$^2$ np$^2$ with n=3, 4 and 5
respectively) the defects belonging to this group feature similar
optical activities (e.g. emission and absorption peaks, singlet
decay lifetimes).\cite{SkujaReview98} A more interesting feature of
the ODCs(II) is that some of their spectroscopic parameters (e.g.
non-radiative decay rates, triplet decay lifetimes) regularly vary
along the isoelectronic series, according to the expected effect of
introducing a heavier central atom along the Si-Ge-Sn series
\cite{SkujaReview98, skuja1992}. Indeed, this property was used as
an important clue to prove a common structural model of the three
defects. Regarding the optical activities, ODC(II) are responsible
of intense signals in the Vis-UV range: Si-ODC(II) mainly gives rise
to a broad nearly-gaussian optical absorption (OA) band centered at
$\sim5.0$ eV due to the transition between the ground singlet
(S$_0$) and the first excited singlet (S$_1$) state. This absorption
excites a fast (lifetime in the ns range) emission band centered at
$\sim4.4$ eV, due to the inverse S$_1$$\rightarrow$S$_0$ transition.
\cite{skuja1994, SkujaReview98} At T$>$300 K, it is possible also to
populate from S$_1$ the first excited triplet state (T$_1$) by the
non-radiative inter-system crossing process (ISC). In these
conditions one observes an additional slow (ms) decaying emission
centered at $\sim2.7$ eV, due to the T$_1$$\rightarrow$S$_0$
transition. The corresponding bands for Ge-ODC(II) are centered at
$\sim5.1$ eV (OA), the fluorescence at $\sim4.3$ eV and
phosphorescence at $\sim3.1$ eV (active at T$>$100
K).\cite{SkujaReview98, agnello2000} Finally, for Sn-ODC(II) they
are centered at $\sim$4.9 eV, $\sim$4.2 eV and $\sim$3.1 eV (active
at T$>$50 K) respectively.\cite{cannizzoSn} The PL signal of ODC(II)
is characteristics of the amorphous phase of silicon dioxide, where
it is usually observed both in as-grown materials or after
irradiation. In contrast it has never been observed in crystalline
SiO$_2$, so that ODC(II) represent an interesting model system to
perform experiments aimed at investigating the peculiar properties
of defects in disordered materials.

Previous optical measurements on the intrinsic and extrinsic ODC(II)
bands have already suggested that ODC(II) are significantly affected
by inhomogeneous effects.\cite{cannizzoSn, leone1, leone2,
cannizzophilos} However, no quantitative estimate exists of their
degree of inhomogeneity. Time-resolved PL measurements reported here
clearly evidence inhomogeneous effects affecting the decay
properties of these defects. Furthermore, they provide for the first
time the possibility of estimating the inhomogeneous and homogeneous
widths of ODC(II) emission bands, and analyze how they depend on the
nature of the central atom.
\section{Experimental Methods}\label{EXP}
We report measurements performed on three samples chosen because
they contain Si-ODC(II), Ge-ODC(II), Sn-ODC(II) defects,
respectively. The first sample is a synthetic silica specimen
(commercial name: Suprasil300), hereafter named S300, with a nominal
concentration of impurities $<$1 ppm in weight. The second one is a
fused silica sample (commercial name: Infrasil301), hereafter named
I301, manufactured by fusion and quenching of natural quartz, with
typical concentration of impurities $\sim$20 ppm in weight. In
particular, as-grown I301 features a $\sim$1 ppm concentration of Ge
impurities, due to contamination of the quartz from which the
material was produced. These two samples were provided by Heraeus
Quartzglas \cite{heraeus} and were $5\times5\times1$ mm$^3$
sized. The last silica sample is doped with 2000 ppm of Sn atoms,
hereafter named Sn-doped silica, prepared by the sol-gel method as
described in reference\cite{chiodini} and rod shaped with a
diameter of 4 mm and thickness of 1.4 mm.

PL measurements were done in a standard back-scattering geometry,
under excitation by a pulsed laser (Vibrant OPOTEK: pulsewidth of 5
ns, repetition rate of 10 Hz, energy density per pulse of
0.30$\pm$0.02 mJ/cm$^2$) tunable in the UV-Visible range. The
luminescence emitted by the sample was dispersed by a spectrograph
(SpectraPro 2300i, PI Acton, 300 mm focal length) equipped by a 300
grooves/mm grating (blaze at 500 nm, bandwidth=3 nm), detected by an
air-cooled intensified CCD (Charge-Coupled Device PIMAX, PI Acton).
The detection system can be triggered so as to acquire the emitted
light only in a given temporal window defined by its width (t$_W$)
and by its delay $t$ from the laser pulse. Our experimental setup
features an acquisition deadtime related to the falling front of
laser pulse. In the following, the origin (t=0) of the time scale is
accordingly defined as the time of the first possible measurement,
corresponding to about $\sim$2ns after the instant of arrival of
laser peak intensity. All PL signals were corrected for spectrograph
dispersion and instrumental response. All measurements reported here
were performed on silica samples kept at cryogenic temperature (10
K) in high vacuum ($\sim10^{-6}$ mbar) within a He flow cryostat
(Optistat CF-V, OXFORD Inst.).

\section{Results}
\begin{figure*}[h!]
\centering \includegraphics[width=8.5 cm]{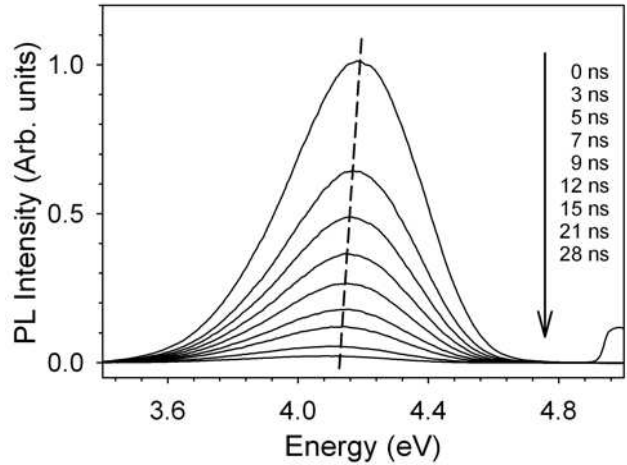}
\caption{Lineshape of Sn-ODC(II) PL activity as excited by 5.00 eV
laser. Different spectra measured at different time delays from
laser pulse are shown. The dashed line follows the PL peak position
for eye-guiding purposes.}\label{fig1}
\end{figure*}

\begin{figure*}[h!]
\centering \includegraphics[width=12 cm]{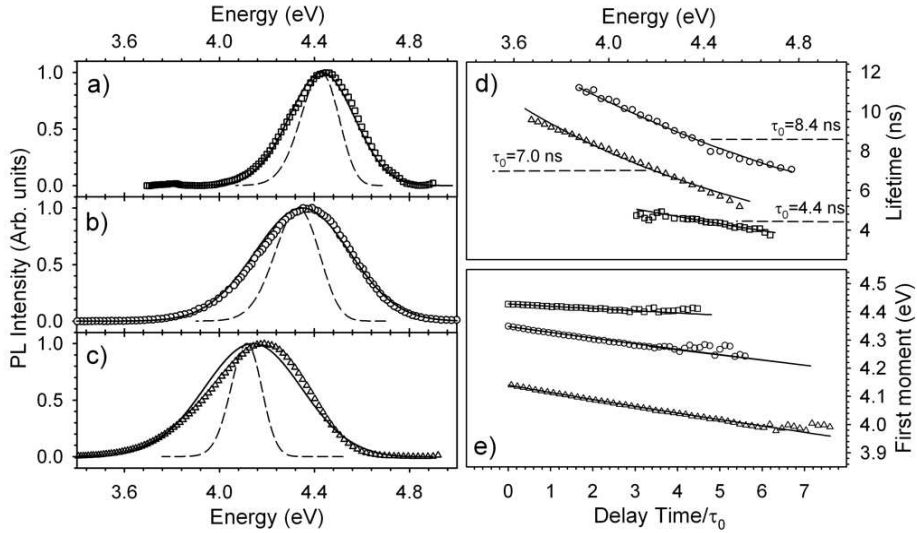} \caption{Low
temperature luminescence lineshape of Si-ODC(II)(Panel-a),
Ge-ODC(II) (Panel-b) and Sn-ODC(II) (Panel-c) at $t$=0. Panel-d:
decay lifetime as measured at different emission energies within the
emission band as measured for Si-ODC(II) (squares), Ge-ODC(II)
(circles) and Sn-ODC(II) (triangles). Panel-e: first moment of the
emission band as a function of time. The continuous lines are the
result of the fitting procedure by the theoretical model in reference
\cite{damicoPRB}; the dashed lines in panels (a)-(c) are the
homogeneous Poissonian line shapes (see discussion).} \label{fig2}
\end{figure*}

In Figure \ref{fig1} we show a typical time-resolved measurement of
the PL activity of Sn-ODC(II) in the Sn-doped silica sample.
The measurement was performed under laser excitation at 248 nm (5.00
eV), corresponding to the peak absorption wavelength of the defect
at cryogenic temperature. The PL signal $L(E,t)$ was monitored by
performing 120 acquisitions with the same integration time $t_W$=0.5
ns but at different delays $t$, going from 0 to 60 ns. The dashed
line reported in Figure \ref{fig1} follows the peak position of PL
band as a function of the delay time from the laser pulse: there is
a clear evidence of a progressive change of the observed PL
lineshape, whose peak moves from 4.2 eV at t=0 ns to 4.1 eV at
t=28 ns. In Figure \ref{fig2}c we report the signal acquired for
$t=0\ ns$, corresponding to the most intense spectrum in Figure
\ref{fig1}. The PL band of Sn-ODC(II), as acquired immediately after
the end of the laser pulse, is peaked at $\sim$4.2 eV and features a
0.48 eV width (Full Width at Half Maximum, FWHM).

Analogous time-resolved measurements were carried out on the PL
activity of Si-ODC(II) defects in the S300 sample using a 248 nm
(5.00 eV) excitation wavelength. The decay was monitored varying $t$
from 0 to 20 ns and with $t_W$=0.5 ns. We report in Figure
\ref{fig2}a the spectrum acquired for $t$=0 ns: the PL band of
Si-ODC(II), as acquired immediately after the end of the laser
pulse, is peaked at $\sim$4.4 eV and features a 0.35 eV FWHM.
Finally, the I301 sample was excited at 240 nm (5.17 eV) and the
luminescence of Ge-ODC(II) was monitored varying $t$ from 0 to 60 ns
with $t_W$=1 ns. As shown in Figure \ref{fig2}b the PL signal
detected in this sample immediately after the end of the laser pulse
is peaked at $\sim$4.3 eV and features a 0.45 eV FWHM.

For each of the three activities (Si-ODC(II), Ge-ODC(II),
Sn-ODC(II)), we calculated the PL decay lifetime $\tau(E)$ at
different emission energies $E$. At the chosen measurement
temperature, the decay kinetics of all the three activities result
to be single-exponential due to quenching of the non-radiative decay
channels.\cite{skuja1994, SkujaReview98, cannizzoSn, agnello2003} As
a consequence, the radiative lifetimes were obtained by fitting with
a single exponential function (\ref{monoexp}) data at several values
of $E$.
\begin{equation}
  I(E,t)= I(E,0)e^{-t/\tau(E)}\label{monoexp}
\end{equation}
In Figure \ref{fig2}d we report so-calculated lifetimes $\tau(E)$ and
we observe that all PL activities in silica feature a dispersion of
the radiative lifetime as a function of the emission energy: the
lifetime goes from $\sim$4 to $\sim$5 ns for the Si-ODC(II), from
$\sim$7 to $\sim$11 ns for the Ge-ODC(II) and from $\sim$5 to
$\sim$9 ns for the Sn-ODC(II). Also, we calculated by numerical
integration from time-resolved spectra $L(E,t)$ (as in Figure
\ref{fig1} for Sn doped sample) the time dependence M$_1$($t$) of
the first moment of the luminescence bands. The temporal behavior of
first moments for the three silica samples are reported in Figure
\ref{fig2}e, where the horizontal axis represents the time delay
$t$ from the laser pulse in units of the central lifetime $\tau_0$
observed at the peak emission energy (indicated in Figure
\ref{fig2}d). We observe that all PL activities in silica feature
an approximately linear decrease of the first moment M$_1$($t$) in
time, with a negative slope increasing with the atomic weight of
ODC(II) defects. The two results are directly connected: indeed, the
dependence of the lifetime from the spectral position within the
emission band corresponds to (and can be alternatively understood
as) a progressive shift of first momentum of PL bands; moreover the
different dependencies of $\tau$ from emission energy of Figure
\ref{fig2}d correspond, as expected, to different slopes in Figure
\ref{fig2}e. Summing up, these experimental findings are the
ultimate reasons that bring about the observed dispersion of the
emission lineshape as observed representatively in Figure \ref{fig1}
for the Sn doped sample.

\section{Discussion}\label{RES}
We have recently proposed a theoretical model that allows to
interpret the behavior of ODC(II) in panels (d) and (e) of Figure
\ref{fig2}, namely the distribution of lifetimes measured at
different emission energies and the correspondent red-shift of the
first moment of the singlet PL band as a function of delay
time.\cite{damicoPRB} The model explains both features in terms of a
gaussian statistical distribution of a single homogeneous parameter
of the defect: the zero phonon (ZPL) energy $E_0$, i.e. the energy
difference between the ground and the first excited electronic
states of the center, both in the ground vibrational sublevel. Every
other spectroscopic feature of the defect (the homogeneous
half-width $\sigma_{ho}$, the oscillator strength $f$, the half Stokes
shift $S$) is assumed as undistributed. Furthermore, the defect is
assumed to be coupled to a single vibrational mode of the matrix
which frequency is thought as the mean frequency of all modes
coupled to the electronic transition. For the present purposes, this
statistical distribution of $E_0$ is equivalent to a distribution of
the peak PL emission energy $E_{e}$,\footnote{E$_e$ is related to
E$_0$ by: E$_e$=E$_0$-$S$. Since $S$ is supposedly undistributed, a
gaussian distribution of E$_0$ is equivalent to a gaussian
distribution of E$_e$, shifted by $S$ and with the same width.} and
physically represents the existence of different environments which
accommodate the defects at different sites of the amorphous matrix. For further details we refer to the original paper.\cite{damicoPRB}

On these basis, one can write a quantitative expression of the PL
emitted by the ensemble of color centers in the amorphous
solid:\cite{damicoPRB}
\begin{eqnarray}
 L_s(E,t|\widehat{E_e},\sigma_{in},\sigma_{ho},f)\propto \nonumber \\
 \int \left[E^3 P(E|E_e,\sigma_{ho}) \cdot e^{-t/\tau(E_e,\sigma_{ho},f)}\right] \cdot
e^{-\frac{\left(E_e-\widehat{E_e}\right)^2}{2\sigma_{in}^2}}dE_e
\label{main}
\end{eqnarray}
Eq. \ref{main} expresses the overall emission signal $L_s(E,t)$
measured at time $t$ at the spectral position $E$. It is obtained by
convolution of a gaussian distribution of $E_e$, whose half-width
$\sigma_{in}$ represents the inhomogeneous linewidth of the ensemble
of defects, with a homogeneous term (within squared parentheses)
representing the PL of a defect emitting at a given value of $E_e$
and with (undistributed) homogeneous half-width $\sigma_{ho}$: the
spectral shape of the homogeneous term is $E^3
P(E|E_{em},\sigma_{ho})$, the $P$ function being a Poissonian with
first moment $E_{e}$ and second moment $\sigma_{ho}$, and its
radiative lifetime is given by the Forster's equation:\cite{forster}
\begin{equation}
 1/\tau(E_e,\sigma_{ho},f)=\frac{2e^2n^2}{m_ec^3\hbar^2}f \int P(E|E_e,\sigma_{ho})E^3 dE \label{ratetau}
\end{equation}
where $f$ is the oscillator strength of the defect, and $n$ is the
refractive index of silica.\footnote{In writing Eq.  \ref{ratetau}
we have neglected the "effective field correction" term, which can
be argued to be close to unity in silica within the investigated
spectral range. We also neglect the slight energy dependence of the
refraction index n which we assume for silica to be 1.5. } Eq.
\ref{ratetau} basically arises from Einstein's relation for
spontaneous emission integrated on the PL lineshape. Summing up, one
can numerically integrate Eq. \ref{main} in order to simulate the
time-resolved PL spectra, $L_s(E,t)$, as a function of the four
parameters $\widehat{E_e},\sigma_{in},\sigma_{ho},f$. From
$L_s(E,t)$ one can then easily calculate the decay lifetime
$\tau_s(E)$ and the kinetics M$_{1s}$($t$) of the first moment, by
using the same procedure applied to the experimental data $L(E,t)$.

Eq. \ref{ratetau} implies that defects with different emission
positions $E_e$ within the gaussian distribution should decay with
different lifetimes. In this sense, it immediately allows to
understand data in Figure \ref{fig2} on a qualitative basis, if we
suppose the PL band of ODC(II) as arising from the inhomogeneous
overlap of bands peaked at different energies, statistically
distributed within the defect population. Also, data in Figure
\ref{fig2} suggest that the degree of inhomogeneity increases while
moving along the Si-Ge-Sn series. In fact, both the slope of
$\tau(E)$ (Figure \ref{fig2}d) and that of M$_1$($t$) (Figure
\ref{fig2}e) grow with increasing atomic weight of the central
atom, thus suggesting the occurrence of progressively stronger
inhomogeneous effects.

This argument can be made quantitative by fitting experimental data,
for each of the investigated PL activities, with our model (Eq.
\ref{main}) in order to estimate $\sigma_{in}$ and $\sigma_{ho}$.
Specifically, we have determined by least-square optimization the
best values of the parameters
($\widehat{E_e},\sigma_{in},\sigma_{ho},f$) that produce a set of
three theoretical curves simultaneously fitting the PL shape at
$t$=0 (Figure \ref{fig2}a, b, c), the dispersion of the decay
lifetimes (Figure \ref{fig2}d) and the kinetics of the first moment
(Figure \ref{fig2}e). The continuous lines in Figure \ref{fig2}
represent the results of our fitting procedure. It is worth noting
the goodness of the fit obtained by using only four free parameters
considering the contemporaneously minimization on spectral and
temporal data. Only for the lineshape at $t$=0 of Sn-ODC(II) the
fitting curve does not reproduce well the experimental data: this
could be due to the presence of another spurious PL signal at lower
energy which apparently enlarges the band, or to the partial failure
for this defect of the approximations inherent in our model, e.g.
more than one homogeneous parameter to be distributed. However, the
behavior of first moment and lifetime dispersion are still well
reproduced by the theoretical model; hence, we consider the
numerical results to be reliable also in this case. In Figure
\ref{fig2}a, b, c we also show, with dashed line, the Poissonian
homogeneous lineshape of half-width $\sigma_{ho}$ as obtained by our
fit procedure for all investigated activities.

\begin{table*}
\caption{Upper section: best fitting parameters obtained by our
theoretical model for the investigated PL activities. Lower section:
values of $\lambda$, $\sigma_{tot}$ (calculated by
$\sigma_{tot}^2=\sigma_{in}^2+\sigma_{ho}^2$), $\hbar \omega_p$ and
$H$, as calculated from best fitting parameters} \label{results}
\begin{center}
\begin{tabular}{c|c|c|c|c}
& $\widehat{E_e}\ [eV]$  & $\sigma_{in}\ [meV]$  & $\sigma_{ho}\ [meV]$  &  $f$ \\
\hline
Si   & 4.42$\pm$0.05  &  110$\pm$10  &  80$\pm$10  &   0.134$\pm$0.019  \\
Ge  & 4.32$\pm$0.05  & 177$\pm$10  &  93$\pm$12  &   0.073$\pm$0.010  \\
Sn   & 4.11$\pm$0.10  &  195$\pm$10  &  65$\pm$10   & 0.104$\pm$0.014 \\
\hline \hline
& $\lambda (\%)$ &   $\sigma_{tot} \ [meV]$  & $\hbar \omega_p\ [meV]$  & $H$ \\
\hline
Si & 65$\pm$4  &  136$\pm$10 &  24$\pm$4 & 11$\pm$4  \\
Ge & 78$\pm$5  &  200$\pm$10 &  23$\pm$6 & 17$\pm$5  \\
Sn  & 90$\pm$5  &  206$\pm$10 & 10$\pm$4 & 40$\pm$10
\end{tabular}\end{center}
\end{table*}

The upper part of Table \ref{results} resumes the best fit
parameters obtained for all investigated PL activities. We calculate
the parameter $\lambda=\sigma_{in}^2/\sigma_{tot}^2$ which estimates
the degree of inhomogeneity. The high values ($>$65\%) of $\lambda$
show that inhomogeneous effects strongly condition the optical
properties of all the ODC(II) defects in silica, $\sigma_{in}$ being
the main contribution to the total width for all the centers. In a
sense, this could be expected \emph{a priori} for a point defect
found exclusively in the amorphous phase of SiO$_2$. On the other
hand, it is worth noting that the degree of inhomogeneity is
remarkably high, particularly for Sn-ODC(II). For this latter
defect, the order of magnitude of the site-to-site fluctuations of
the emission peak position results to be as large as
$\sigma_{in}$$\sim$0.2 eV. The degree of inhomogeneity
systematically varies along the isoelectronic series: the value of
$\lambda$ of extrinsic Sn-ODC(II) defects (Sn-doped sample) is
higher than that of Ge-ODC(II) extrinsic centers (I301 sample),
which is in turn higher than Si-ODC(II) intrinsic defects (S300
sample). These variations of $\lambda$ are mainly due to the growth
of $\sigma_{in}$ with atomic weight, since variations of
$\sigma_{ho}$ are weaker. This trend can be tentatively interpreted
as follows: Ge and Sn impurities are isoelectronic to Si atoms and
thus able to be accommodated in substitutional positions.
Nonetheless, the distortion they cause to the matrix presumably
extends over a larger surrounding volume than a single SiO$_2$
tetrahedra due to their being bigger and heavier than their
intrinsic counterpart. A bigger volume affected by the presence of
the defect is expected to result in a higher sensitivity to
site-to-site structural fluctuations, which eventually causes
stronger fluctuations of $E_e$. These considerations based on the
experimental results reported here are at variance with previous
ones based on computational findings in reference\cite{pacchioni} where it was argued that heavier Ge and Sn
atoms are much less sensitive to the details of local geometry. In
this regard, we want to stress that the phonon-assisted ISC process
at room temperature, responsible of excitation of the triplet band
in ODCs(II), is more efficient for heavier atoms in the
isoelectronic series Si-Ge-Sn:\cite{skuja1992} this indicates a
stronger coupling with environment for impurities atoms. The results
reported here complete the characterization of the isoelectronic
series of oxygen deficient centers in silica, by yielding
information about their inhomogeneous properties, which adds to
existing knowledge founded on traditional spectroscopic
investigation.

Other two parameters of interest can be calculated from
$\sigma_{ho}$: the vibrational frequency $\hbar \omega_p
=\sigma_{ho}^2/S$ and the Huang-Rhys factor $H=S^2/\sigma_{ho}^2$.
In these expressions, the parameter $S$ represents the half Stokes
shift, estimated experimentally by measuring the half-difference
between the spectral positions of the excitation energy and emission
peaks. $S$ results to be: 0.27 eV, 0.38 eV and 0.41 eV, in Si, Ge
and Sn-ODC(II) respectively. Based on these values of $S$, we
calculate $\hbar \omega_p$ and $H$, reported in the lower part of
Table \ref{results}. The vibrational frequencies found here for
ODC(II) defects show that all of them are preferentially coupled
with very low frequency vibrational modes, accordingly with previous
experimental and computational results.\cite{cannizzoSn, galeener,
umari} Albeit the relatively high uncertainty on $\hbar \omega_p$ as
determined by the fitting procedure, data show a decreasing trend
while going from the lightest to the heavier ODC(II). Qualitatively,
this is to be expected if one roughly assumes that the variations in
the force constant of the vibration are negligible from Si to Sn:
indeed, the frequency of a mode highly localized on the central atom
should in this case be inversely proportional to the square of its
mass. It is possible to object that the lower  $\hbar \omega_p$
frequency obtained for Sn-ODC(II) could be affected by the worse
fitting result obtained on its PL lineshape; on the other side, it
is worth stressing that the above discussed results on heterogeneity
are poorly affected by this fact, as they mainly depend on the well
fitted slope of the first momentum and lifetime dispersion curves.

Finally, the values of the oscillator strength found here, are in
excellent agreement with those reported in a review paper about
oxygen deficiency centers in silica \cite{SkujaReview98}: 0.03-0.07
for Ge-ODC(II) and 0.15 for Si-ODC(II).

\section{Conclusions}\label{CONCL}
We studied by time-resolved luminescence the defects belonging to
the isoelectronic series of oxygen deficient centers in amorphous
silicon dioxide. The dispersion of the emission lineshape was used
as a probe to quantitatively evaluate the influence of inhomogeneous
effects on the optical properties of the defects. We provided for
Si-ODC(II), Ge-ODC(II) and Sn-ODC(II) an estimate of the
inhomogeneous and homogeneous widths, on the grounds of a simple
theoretical model that is able to well reproduce all experimental
findings. The degree of inhomogeneity of the defects turns out to
grow regularly with the atomic weight of the central atom, while the
variations of the homogeneous properties are weaker. Along with the
homogeneous width, we estimated also the other homogeneous
parameters of oxygen deficient centers: oscillator strength,
Huangh-Rhys factor and vibrational frequency of the electron-phonon
interaction.
\paragraph{Acknowledgment}
We acknowledge financial support received from project "P.O.R.
Regione Sicilia - Misura 3.15 - Sottoazione C". The authors would
like to thank G. Lapis and G. Napoli for assistance in cryogenic
work. Finally we are grateful to LAMP research group
(http://www.fisica.unipa.it/amorphous/) for support and enlightening discussions.

\end{document}